\documentclass[12pt,a4paper]{article}

\newcommand{\balpha}{\mbox{\boldmath$\alpha$}}
\newcommand{\bdelta}{\mbox{\boldmath$\delta$}}
\newcommand{\beeta}{\mbox{\boldmath$\eta$}}
\newcommand{\blambda}{\mbox{\boldmath$\lambda$}}
\newcommand{\bxi}{\mbox{\boldmath$\xi$}}
\newcommand{\bzeta}{\mbox{\boldmath$\zeta$}}

\usepackage{latexsym}
\usepackage{wasysym}

\title{Isometry germs and related structures}

\author{Ll.\ Bel \footnote{wtpbedil@lg.ehu.es}\\
\emph{Fisika Teorikoa, Euskal Herriko Unibertsitatea}, \\
\emph{P.K. 644, 48080 Bilbo, Spain}
}

\begin{document}
\maketitle

\begin{abstract}

We define an Isometry germ at any given event $x$ of space-time as a vector field $\bxi$ defined in a neighborhood of $x$ such that the Lie derivative of both the metric and the Riemannian connection are zero at this event. Two isometry germs can be said to be equivalent if their values and the values of their first derivatives coincide at $x$. The corresponding quotient space can be endowed with a structure of a bracket algebra which is a deformation of de Sitter's Lie algebra. Each isometry germ defines also a local stationary frame of reference, the consideration of the family of adapted coordinate transformations between any two of them leading to a local novel structure that generalizes the Lorentz group. 

\end{abstract}

\section*{Introduction}

Being able to model space-times both at the local and global scales is one of the glories of General Relativity. But in both domains obscure points and unsettled problems remain. This paper deals with one of these problems  at a local scale. We would like to know how to describe the  the physical environment in the neighborhood $\cal U$ of an event $x$ of space-time when we only know at this event the space-time curvature, represented by the Riemann tensor. It is often said that if $\cal U$ is small enough compared to some characteristic length derived from the Riemann tensor then we can forget about General Relativity and deal with $\cal U$ as if were a small domain of Minkowski's space-time. We want to establish the formulas that will allow us to go beyond this intuition and decide when further work is required. 

Two particular domains might benefit from an improvement of our knowledge of local gravitational fields. They both rely heavily on the ability  to refer to the Poincar\'{e} and Lorentz groups. Consider the following problems: ¿How to define particles if we can not associate them with irreducible unitary representations of the Poincar\'{e} group when we know that this is not a symmetry group at an appropriate approximation? Or: ¿Which is the Principle of correspondence on which to base Quantum Mechanics if we can not rely on the existence of privileged coordinates as are the Cartesian coordinates for classical and Special Relativity?.

Another related problem is that of defining a sufficiently large class of frames of reference with the same number of degrees of freedom as that of rigid frames of reference in classical mechanics. Consider the following problem: Two observers $\bf O_1$ and $\bf O_2$ located at a given point of space have measured their relative velocity and each of them has measured its own Newtonian $\bf \Lambda_{1,2}$ and Coriolis $\bf \Omega_{1,2}$ fields. Let us assume that in the frame of reference of $\bf O_1$ there is a local constant electric field. ¿What is the electromagnetic field in the frame of reference of $\bf O_2$? If the Riemann tensor were zero and the two observers were inertial observers, i.e. if $\Lambda_{1,2}=0$ and $\Omega_{1,2}=0$, then a Lorentz transformation will give us the answer. ¿What do we have to do in the general case?

These are the problems and the questions that have motivated this paper. In the process of our work  we have found a Bracket algebra $A$ which is a deformation of the Lie algebra of de Sitter's group, but $A$ is not a Lie Algebra and therefore it does not generate a group. We have found also a family of transformations that is a deformation of the Lorentz group and it is not a group. It is instead a novel structure whose general interest remains to be established. This paper then does not accomplish very much. The question is: is it the first step in the right direction ? 

\section{Isometry germs and de Sitter's deformed algebra}

Let us consider a vector field $\bxi$ defined on a neighborhood $\cal U$ of an event $x$. We shall say that $\bxi$ is an isometry germ at $x$ if both the Lie derivative (e.g. \cite{Lichnerowicz}, \cite{Yano})\,\footnote{All our references are indicative and none of them is exclusive of many other possible ones.} of the metric and the Lie derivative of the Riemannian connection are zero at the event $x$, i.e. if:

\begin{equation}
\label{1.1}
{\cal L}(\xi)g_{\alpha\beta}|_x=0, \quad
{\cal L}(\xi)\Gamma^\mu_{\alpha\beta}|_x=0
\end{equation}
Or, equivalently, if:

\begin{equation}
\label{1.2}
\nabla_\alpha\xi_\beta+\nabla_\beta\xi_\alpha\sim 0, \quad
\nabla_\alpha\nabla_\beta\xi^\gamma+R^\gamma_{\beta\rho\alpha}\xi^\rho\sim 0
\end{equation}
where {\bfseries R} is the Riemann tensor and where from now on we shall use the symbol $\sim$ to mean an equality at a particular event $x$. It follows from these equations that:

\begin{equation}
\label{1.3}
\nabla_\alpha\xi^\beta\sim\zeta_{\ \alpha}^\beta \quad\hbox{with}\quad
\zeta_{\alpha\beta}=-\zeta_{\beta\alpha} 
\end{equation}
and:

\begin{equation}
\label{1.4}
\nabla_\alpha\nabla_\beta\xi^\gamma\sim -R^\gamma_{\ \beta\rho\alpha}\lambda^\rho
\end{equation}
where $\blambda$ is the restriction of $\bxi$ at $x$.

We shall say that two isometry germs are equivalent if their restrictions and the restrictions of their first and second partial derivatives coincide at $x$. Every class of equivalence, that we shall call also an isometry germ, is characterized by a pair $a:(\blambda, \bzeta)$. Since the sum of two isometry germs is an isometry germ and the product of a number with an isometry germ is again an isometry germ it follows that the space of germs $A$ is isomorphic with the direct sum of the tangent space at $x$ and its exterior square: $A=T_x\oplus\Lambda^2_x$ 

We shall use the following notations:

\begin{equation}
\label{1.5}
\Re a= \blambda, \quad \Im a=\bzeta
\end{equation}

Let us consider two isometry germs $a_1:(\bf\lambda_1,\bf\zeta_1)$, $a_2:(\bf\lambda_2,\bf\zeta_2)$ $\in A$ and let $\bxi_1$ and $\bxi_2$ be any two particular members of the corresponding classes of equivalence. The Lie bracket (e.g. \cite{Lichnerowicz}) $\bxi_3$ of these vectors fields is by definition:

\begin{equation}
\label{1.6}
\xi_3^\alpha=[\bxi_1, \bxi_2]^\alpha=\xi_1^\rho\nabla_\rho\xi_2^\alpha-\xi_2^\rho\nabla_\rho\xi_1^\alpha
\end{equation}  
We define the bracket of $a_1$ and $a_2$:

\begin{equation}
\label{1.6.0}
a_3=[a_1,a_2]
\end{equation}
as the element $a_3:(\blambda_3, \bzeta_3)$ of $A$ such that:

\begin{equation}
\label{1.6.1}
\lambda_3^\alpha\sim\xi_3^\alpha, \quad 
\zeta_{3\beta}^{\alpha}\sim\nabla_\beta\xi^\alpha
\end{equation} 
The element $a_3$ thus defined does depend only on the elements $a_1$ and $a_2$ and not on the particular elements $\bxi_1$ and $\bxi_2$ chosen as representatives of the corresponding classes. This is obvious from the simple calculation below. 

From (\ref{1.6})it follows that:

\begin{equation}
\label{1.7}
\lambda_3^\alpha=\lambda_1^\rho\zeta_{2\rho}^\alpha-
\lambda_2^\rho\zeta_{1\rho}^\alpha
\end{equation}
Also covariantly deriving (\ref{1.6}) and using (\ref{1.3}), (\ref{1.4}) and the symmetries of the Riemann tensor we obtain:

\begin{equation}
\label{1.8}
\zeta_{3\beta}^\alpha=
\zeta_{1\beta}^\rho\zeta_{2\rho}^\alpha
-\zeta_{2\beta}^\rho\zeta_{1\rho}^\alpha
+R^\alpha_{\ \beta\rho\sigma}\lambda_1^\rho\lambda_2^\sigma
\end{equation}

Let us consider a local system of coordinates around the event $x$ such that:

\begin{equation}
\label{1.8.1}
g_{\alpha\beta}\sim \eta_{\alpha\beta}
\end{equation}
{\bfseries g} being the space-time metric and {\beeta} being the Cartesian form of the Minkowski metric so that, {\bfseries e} being the induced natural basis of $T_x$, we have:

\begin{equation}
\label{1.8.2}
e_\alpha e_\beta =\eta_{\alpha\beta}
\end{equation}
From now on raising and lowering greek indices will be done with this latter metric. Also from now on
we shall use the following basis ($P_\alpha, J_{\mu\nu}$), with $\mu<\nu$, of $T_x\oplus\Lambda^2_x$ defined by:

\begin{equation}
\label{1.9}
(\Re P_\alpha)^\beta=\delta_\alpha^\beta, \quad (\Im P_\alpha)_\nu^\mu=0
\end{equation}
and:

\begin{equation}
\label{1.10}
(\Re J_{\mu\nu})^\alpha=0, \quad
(\Im J_{\mu\nu})^\rho_{\ \sigma}=\eta_{\sigma\tau}\delta_{\mu\nu}^{\rho\tau} 
\end{equation}
where the $\bdelta$'s are  respectively the Kronecker tensors of order 2 and 4. Notice that although the restriction $\mu<\nu$ is necessary to have the right number of elements in the basis we in the sequel shall not consider this restriction and we shall instead insert when necessary the appropriate numerical factors to keep track of the skew symmetry between these two indices. 

To keep notations simple we shall use the same notation $P_\alpha, J_{\mu\nu}$ to refer to the isometry germs that they generate. From this point of view the defining conditions (\ref{1.9}) and (\ref{1.10}) mean:

\begin{equation}
\label{1.10.1}
P_\alpha^\beta\sim\delta_\alpha^\beta, \quad \nabla_\rho P_\alpha^\beta \sim 0 
\end{equation}
and:

\begin{equation}
\label{1.10.2}
J_{\mu\nu}^\alpha \sim 0, \quad  \nabla_\alpha J_{\mu\nu}^\sigma\sim\eta_{\alpha\rho}\delta_{\mu\nu}^{\sigma\rho} 
\end{equation}
Using Eqs. (\ref{1.7})and (\ref{1.8}) we obtain to begin with the following commutation relations:

\begin{equation}
\label{1.11}
(\Re[P_\alpha,P_\beta])^\rho=0, 
\quad (\Im[P_\alpha,P_\beta])^\rho_{\ \sigma}
=\frac12 R^\rho_{\ \sigma\alpha\beta} 
\end{equation}
that are equivalent to:

\begin{equation}
\label{1.11.1}
[P_\alpha,P_\beta]
=\frac12 R_{\alpha\beta}^{\ \ \mu\nu}J_{\mu\nu} 
\end{equation}
the remaining commutation relations being the familiar ones:  
\begin{equation}
\label{1.12}
[J_{\lambda\mu},P_\alpha]=(\eta_{\lambda\alpha}\delta^\rho_\mu-
\eta_{\mu\alpha}\delta^\rho_\lambda)P_\rho 
\end{equation}

\begin{equation}
\label{1.13}
[J_{\alpha\beta},J_{\lambda\mu}]=(
\eta_{\alpha\lambda}\delta_{\beta\mu}^{\rho\sigma}+
\eta_{\beta\mu}\delta_{\alpha\lambda}^{\rho\sigma}-
\eta_{\alpha\mu}\delta_{\beta\lambda}^{\rho\sigma}-
\eta_{\beta\lambda}\delta_{\alpha\mu}^{\rho\sigma})J_{\rho\sigma}
\end{equation}

If the space-time has constant curvature $K$, i.e. if (e.g. \cite{Eisenhart}):

\begin{equation}
\label{1.14}
R_{\alpha\beta}^{\ \ \rho\sigma}=\frac{R}{12}\delta_{\alpha\beta}^{\rho\sigma}
\end{equation}
where $K=R/12$ and $R$ is the Ricci scalar, then Eqs. (\ref{1.11}) become (\cite{Gursey}):

\begin{equation}
\label{1.15}
[P_\alpha,P_\beta]=\frac{R}{12}J_{\alpha\beta} 
\end{equation}
and the bracket algebra $A$ that we have constructed is the Lie algebra of de Sitter's group, or the Poincar\'{e} group if $R=0$. But in general $A$ is not a Lie algebra and as we prove below is a Lie algebra {\bfseries iff} the space-time at the event $x$ has constant curvature.

The algebra $A$ is a Lie algebra {\bfseries iff} the Jacobi identities:

\begin{equation}
\label{1.16}
[[X_I,X_J],X_K]+[[X_J,X_K],X_I]+[[X_K,X_I],X_J]=0,  
\end{equation}
where $X_I$, $X_J$ and $X_K$ are three arbitrary elements of a basis of $A$, are satisfied. 

We know that the Jacobi expressions above when they involve none or only one of the generators $\bf P$'s are always satisfied because their calculation, being identical to that with a null Riemann tensor,  gives zero. On the other hand if we consider the case with three generators $\bf P$'s the result is again zero because of the cyclic symmetry of the Riemann tensor.  

There remains to consider the case with $X=P_\alpha$, $Y=P_\beta$ and $Z=J_{\lambda\mu}$ then (\ref{1.16}) that leads to:

\begin{eqnarray}
\label{1.18}
\nonumber R_{\ \alpha\beta\mu\ }^\sigma\delta^\rho_\lambda
+R_{\ \alpha\beta\lambda\ }^\rho\delta^\sigma_\mu
-R_{\ \alpha\beta\mu\ }^\sigma\delta^\sigma_\lambda 
-R_{\ \alpha\beta\lambda\ }^\sigma\delta^\rho_\mu \\ 
+R_{\lambda\alpha\ \ }^{\ \ \rho\sigma}\eta_{\beta\mu}
+R_{\mu\beta\ \ }^{\ \ \rho\sigma}\eta_{\alpha\lambda}
-R_{\mu\alpha\ \ }^{\ \ \rho\sigma}\eta_{\beta\lambda}
-R_{\lambda\beta\ \ }^{\ \ \rho\sigma}\eta_{\alpha\mu}=0
\end{eqnarray}
Contraction of the indices $\beta$ and $\mu$ leads to:

\begin{eqnarray}
\label{1.19}
-R^\sigma_\alpha\delta^\rho_\lambda+R^\rho_\alpha\delta^\sigma_\lambda
-R^{\ \rho\ \sigma}_{\alpha\ \lambda\ }+R^{\ \sigma\ \rho}_{\alpha\ \lambda\ }
+2R^{\ \ \rho\sigma}_{\lambda\alpha\ \ }=0
\end{eqnarray}
where the two indices $R$ is the Ricci tensor. A further contraction of $\alpha$ and $\sigma$ leads to:

\begin{equation}
\label{1.20}
R^\rho_\lambda=\frac14 R\delta^\rho_\lambda 
\end{equation}
Using now this formula in (\ref{1.19}), remembering that $R=12K$ we obtain (\ref{1.14}).

Notice that this result was easy to guess from the knowledge of the theorem stating that a space-time has a symmetry group with maximal dimension $10$ {\bfseries iff} it has constant curvature, but it is still interesting to have a simpler pure algebraical proof in an enlarged context.

\section{The Casimir operator}

The capital indices $I,J,K,\cdots$ that we have already used in the preceding section stand either for a single covariant index $\alpha,\beta,\gamma,\cdots$ or for an antisymmetric pair $\rho\sigma,\mu\nu\cdots$. Using this notation we  write generically the commutation relations (\ref{1.11.1})-(\ref{1.13})  as:

\begin{equation}
\label{2.1}
[X_I,X_J]=C_{IJ}^KX_K
\end{equation}
where the non zero structure constants {\bfseries C}'s are:

\begin{eqnarray}
\label{2.2}
C^{\lambda\mu}_{\alpha,\beta}&=&R_{\alpha\beta\ \ }^{\ \ \lambda\mu}\\
C^\rho_{\lambda\mu,\alpha}&=&
\eta_{\lambda\alpha}\delta_\mu^\rho-
\eta_{\mu\alpha}\delta_\lambda^\rho \\
C^{\rho\sigma}_{\alpha\beta,\lambda\mu}&=&
\eta_{\alpha\lambda}\delta_{\beta\mu}^{\rho\sigma}+
\eta_{\beta\mu}\delta_{\alpha\lambda}^{\rho\sigma}-
\eta_{\alpha\mu}\delta_{\beta\lambda}^{\rho\sigma}-
\eta_{\beta\lambda}\delta_{\alpha\mu}^{\rho\sigma}
\end{eqnarray}
Notice the use of a comma to separate single covariant indexes from covariant pairs ones.

Although we are not dealing in general with a Lie algebra some of the algebraic objects which are relevant in the study of Lie algebras can also be defined for the deformed algebra we are considering here. This is the case with the Killing tensor and the first Casimir operator (e.g. \cite{Dixmier}, \cite{Hammermesh}).

The Killing tensor will thus be defined as the symmetric tensor with components:

\begin{equation}
\label{2.3}
K_{IJ}=BC^K_{IL}C^L_{JK}
\end{equation}
where $B$ is a free factor that we shall choose in a moment. A simple calculation gives:

\begin{eqnarray}
\label{2.4}
K_{\rho,\sigma}&=&-2BR_{\rho\sigma}\\
K_{\alpha\beta,\gamma\delta}&=&-6B(\eta_{\alpha\gamma}\eta_{\beta\delta}
-\eta_{\alpha\delta}\eta_{\beta\gamma})\\
K_{\alpha\beta,\gamma}&=&0
\end{eqnarray}

The determinant of the Killing tensor is different of zero {\bfseries iff} the determinant of the Ricci tensor is different of zero. Let us assume that this is the case. We choose then $B$ to be, up to a sign $\epsilon$: 

\begin{equation}
\label{2.4.1}
B=-\frac{\epsilon}{2}|\det(R_{\mu\nu})|^{-1/4}
\end{equation}
which is a convenient factor to use below when considering Einstein's space-times.
Let us consider the inverse of the Killing tensor and that of the Ricci tensor:

\begin{equation}
\label{2.6}
K_{IS}K^{SJ}=\delta^J_I, \quad R_{\alpha\rho}U^{\rho\beta}=\delta^\beta_\alpha
\end{equation} 
We have more explicitly:

\begin{equation}
\label{2.6.1}
K^{\rho,\sigma}=-\frac{1}{2B}U^{\rho\sigma}, \quad
K^{\gamma\delta,\rho\sigma}=-\frac{1}{6B}(\eta^{\gamma\rho}\eta^{\delta\sigma}-
\eta^{\gamma\sigma}\eta^{\delta\rho})
\end{equation}

Let us assume now that a representation of $A$ has been found. If we designate the operators of the representation algebra generically as $\widehat{X_I}$, i.e. if:

\begin{equation}
\label{2.7}
[\widehat{X}_I,\widehat{X}_J]=\widehat{X}_I\widehat{X}_J-\widehat{X}_J\widehat{X}_I=C^K_{IJ}\widehat{X}_K 
\end{equation} 
then the Casimir operator is:
\begin{equation}
\label{2.8}
\widehat{C}= K^{IJ}\widehat{X}_I\widehat{X}_J 
\end{equation}
or more explicitly:

\begin{equation}
\label{2.9}
\widehat{C}=K^{\rho\sigma}\widehat{P}_\rho\widehat{P}_\sigma
-\frac{1}{12B}\eta^{\alpha\rho}\eta^{\beta\sigma}\widehat{J}_{\alpha\beta}\widehat{J}_{\rho\sigma}
\end{equation}
where we have dropped the by now useless comma to separate the contravariant indices of the symmetric tensor $\bf K$.
If the space-time is an Einstein space at the event $x$, i.e. if:

\begin{equation}
\label{2.10}
R_{\alpha\beta}\sim \frac{R}{4}\eta_{\alpha\beta}, \quad U^{\alpha\beta}=\frac{4}{R}\eta^{\alpha\beta}
\end{equation}
then, even without the space-time having constant curvature, 
the Casimir operator becomes, using the numerical factor $B=-2/R$ given in this case by (\ref{2.4.1}) with $\epsilon=1$ (e.g. \cite{Gursey}):

\begin{equation}
\label{2.11}
\widehat{C}=\eta^{\rho\sigma}\widehat{P}_\rho\widehat{P}_\sigma
+\frac{R}{24}\eta^{\alpha\rho}\eta^{\beta\sigma}\widehat{J}_{\alpha\beta}
\widehat{J}_{\rho\sigma}
\end{equation}
which is the usual expression for it when the space-time has constant curvature. 
Notice that this expression has the expected limit when $R$ tends to zero while the whole process is invalid if one assumes that $R=0$ from the beginning. 	

The importance of the Casimir operator of Lie algebras comes from the fact that it commutes with all the elements of the representation algebra, i.e. :

\begin{equation}
\label{2.11.1}
[\widehat{C},\widehat{X}_I]=0.
\end{equation} 
But the algebra $A$ that we are studying is not a Lie algebra and the Eqs. above are not satisfied in general as we show below.

Let us consider first the commutation relations:

\begin{equation}
\label{2.12}
[\widehat{J}_{\alpha\beta},\widehat{C}]= K^{\rho\sigma}[\widehat{J}_{\alpha\beta},\widehat{P}_\rho\widehat{P}_\sigma]-\frac{1}{12B}[\widehat{J}_{\alpha\beta},\eta^{\mu\rho}\eta^{\nu\sigma}\widehat{J}_{\mu\nu}\widehat{J}_{\rho\sigma}]
\end{equation}
The second bracket in the r-h-s does not involve the Riemann tensor and it is known to be zero. As for the first bracket, a simple calculation with a repeated use of the formula:

\begin{equation}
\label{2.12.1}
[\widehat{X}_I,\widehat{X}_J\widehat{X}_K]=
\widehat{X}_I[\widehat{X}_J,\widehat{X}_K]+
[\widehat{X}_I,\widehat{X}_J]\widehat{X}_K
\end{equation}
leads to:

\begin{equation}
\label{2.13}
[\widehat{J}_{\alpha\beta},\widehat{C}]= -K^\rho_\alpha(P_\rho P_\beta+P_\beta P_\rho)+K^\rho_\beta(P_\rho P_\alpha+P_\alpha P_\rho)
\end{equation}
The r-h-s is not zero in general but it is zero trivially if the space-time is an Einstein space. 

Let us consider now the commutation relations:

\begin{equation}
\label{2.14}
[\widehat{P}_\alpha,\widehat{C}]= K^{\rho\sigma}[\widehat{P}_\alpha,\widehat{P}_\rho\widehat{P}_\sigma]-\frac{1}{12B}[\widehat{P}_\alpha,\eta^{\mu\rho}\eta^{\nu\sigma}\widehat{J}_{\mu\nu}\widehat{J}_{\rho\sigma}]
\end{equation}
We obtain then:

\begin{equation}
\label{2.15}
[\widehat{P}_\alpha,\widehat{C}]= \frac12 K^{\rho\sigma}
R_{\alpha\sigma\ \ }^{\ \ \gamma\delta}(\widehat{P}_\rho\widehat{J}_{\gamma\delta}+
\widehat{J}_{\gamma\delta}\widehat{P}_\rho)
-\frac{1}{12B}\eta^{\rho\sigma}(\widehat{J}_{\rho\alpha}\widehat{P}_\sigma+\widehat{P}_\sigma\widehat{J}_{\rho\alpha})
\end{equation}
This commutator is zero {\bfseries iff} the space-time has constant curvature at the event $x$. That is to say that it is not zero even in the case where the space-time is an Einstein space, case for which the Casimir operator is formally identical to the Casimir operator of an space-time with constant curvature.  

\section{The Laplacian operator}

The natural principle of correspondence between vector fields $\bxi$ and differential operators:

\begin{equation}
\label{0.1}
\bxi\rightarrow \hat\xi=\xi^\rho\nabla_\rho
\end{equation}
can not be used to obtain an intrinsic representation of the algebra $A$ because the third derivatives of the isometry germs are not defined except when the space-time has constant curvature. It can be useful though sometimes as a debased representation if we keep in mind that only the value of $\hat\xi$ and its first and second derivatives are defined at the event $x$, while only the values of:

\begin{equation}
\label{0.2}
\hat\xi_1\hat\xi_2=\xi_1^\rho\nabla_\rho(\xi_2^\sigma\nabla_\sigma)
\end{equation}
and its first derivatives are defined at $x$, and while only the value of:
\begin{equation}
\label{0.3}
\hat\xi_1\hat\xi_2\hat\xi_3=
\xi_1^\rho\nabla_\rho(\xi_2^\sigma\nabla_\sigma(\xi_3^\tau\nabla_\tau)
\end{equation}
is defined at the event $x$.

We give in the following one example of its usefulness.
As we mention before the Lie algebra of the Poincar\'{e} group, i.e. the algebra $A$ when the Riemann tensor is zero, does not have a Casimir operator in the sense of the definition (\ref{2.8}) because the Ricci tensor being zero the tensor {\bfseries U} does not exists. Nevertheless the operator:

\begin{equation}
\label{3.1}
\widehat C=\eta^{\rho\sigma}\widehat{P}_\rho\widehat{P}_\sigma
\end{equation}
can be defined as the limit of the Casimir operator (\ref{2.11}) when the Ricci scalar $R$ goes to zero. Although it is most often defined as the lowest degree element of the envelopping algebra of $A$ which commutes with all the elements of $A$ (e.g. \cite{Okubo}). Another fundamental property for physics is that using the representation:

\begin{equation}
\label{3.2}
\widehat{P}_\rho=P_\rho^\sigma\partial_\sigma
\end{equation}
$\widehat C$ becomes the Laplacian operator :

\begin{equation}
\label{3.3}
\widehat C=\eta^{\rho\sigma}\partial_{\rho\sigma}
\end{equation}
a property that does not even has in general the restriction, at the event $x$, of the Casimir operator defined by Eq. (\ref{2.8}).

We give below the definition of a new general operator:

\begin{equation}
\label{3.4}
{\widehat{C}}_\triangle=\eta^{\rho\sigma}\widehat{P}_\rho\widehat{P}_\sigma
+\frac12\eta^{\alpha\rho}(R^{\beta\sigma}-\frac16 R\eta^{\beta\sigma})\widehat{J}_{\alpha\beta}\widehat{J}_{\rho\sigma}
\end{equation}
that although it has not been derived from a general procedure it has two important properties: i) it reduces obviously to the Casimir operator (\ref{2.11})
when the space-time has constant curvature or is an Einstein space, and reduces to (\ref{3.1}) when it is flat, and ii) it is osculator to the Laplacian operator. We mean by this that if we set : 

\begin{equation}
\label{3.5}
\widehat{P}_\rho=P_\rho^\sigma\nabla_\sigma, \quad \widehat{J}_{\rho\sigma}=J_{\rho\sigma}^\alpha\nabla_\alpha
\end{equation}
then the following relations can be proved easily using (\ref{1.10.1}) and (\ref{1.10.2}):

\begin{equation}
\label{3.6}
{\widehat{C}}_\triangle\sim \triangle, \quad \nabla_\alpha{\widehat{C}}_\triangle\sim\nabla_\alpha\triangle.
\end{equation}
where:

\begin{equation}
\label{3.7}
\triangle=g^{\alpha\beta}\nabla_\alpha\partial_\beta
\end{equation}
So that we can consider ${\widehat{C}}_\triangle$ to be an algebraic substitute for the Laplacian operator when dealing with linear representations of $A$. 

The commutation relations of this new operator with the generators of the algebra $A$ are now the following:

\begin{eqnarray}
\label{3.8}
[\widehat{P}_\alpha,\widehat{C}]=\frac12 R_{\alpha\rho}^{\ \ \gamma\delta}\eta^{\rho\sigma}(\widehat{P}_\sigma\widehat{J}_{\gamma\delta}+
\widehat{J}_{\gamma\delta}\widehat{P}_\sigma) \nonumber \\
+\frac12(R^{\mu\nu}-\frac16 R\eta^{\mu\nu})\eta^{\rho\sigma}
(\widehat{P}_\rho\widehat{J}_{\sigma\nu}+
\widehat{J}_{\sigma\nu}\widehat{P}_\rho)
-\widehat{P}_\mu\widehat{J}_{\alpha\nu}-
\widehat{J}_{\alpha\nu}\widehat{P}_\mu)
\end{eqnarray}
and:

\begin{eqnarray}
\label{3.9}
[\widehat{J}_{\alpha\beta},\widehat{C}]=\frac12(R^\mu_\beta-\frac16 R\delta^\mu_\beta)\eta^{\rho\nu}(\widehat{J}_{\rho\mu}\widehat{J}_{\alpha\nu}+
\widehat{J}_{\alpha\nu}\widehat{J}_{\rho\mu}) \nonumber \\
-\frac12(R^\mu_\alpha-\frac16 R\delta^\mu_\alpha)\eta^{\rho\nu}(\widehat{J}_{\rho\mu}\widehat{J}_{\beta\nu}+
\widehat{J}_{\beta\nu}\widehat{J}_{\rho\mu})
\end{eqnarray}

For space-times with constant curvature the Principle of correspondence (\ref{0.1}) is a representation of de Sitter's algebra and the Casimir operator (\ref{2.11}) coincides also with the Laplacian of the space-time\,\footnote{J.\ Mart\'{\i}n, private communication}

\section{Adapted local coordinate systems}

An obvious system of coordinates to use to deepen the geometric and physical understanding of the family of symmetry germs generated by the algebra $A$ of the preceding sections is any system of geodesic coordinates {\bfseries z} anchored at an event $x$ with a natural orthonormal basis at $x$. With such a system of coordinates we have (e.g. \cite{Thomas}):

\begin{equation}
\label{4.1}
g_{\alpha\beta}\sim \eta_{\alpha\beta}, \quad \Gamma^\gamma_{\alpha\beta}\sim 0,
\quad \partial_\lambda \Gamma^\gamma_{\alpha\beta}\sim 
\frac13(R^\gamma_{\ \alpha\lambda\beta}+R_{\ \beta\lambda\alpha})
\end{equation}
from where it follows that the space-time metric can be written up to the second order of approximation as:

\begin{equation}
\label{4.2}
g_{\alpha\beta}\approx\eta_{\alpha\beta} 
-\frac13 R_{\alpha\lambda\beta\mu}z^\lambda z^\mu
\end{equation}
where from now on the symbol $\approx$ will mean that products of more than two coordinate values have been neglected unless more terms are written explicitly.
Let $\bxi$ be an isometry germ generated by the element $a: (\blambda, \bzeta)$. From (\ref{1.3}) and (\ref{1.4}) we get then:

\begin{equation}
\label{4.3}
\xi^\gamma\approx\lambda^\gamma+z^\rho\zeta_{\ \rho}^\gamma
-\frac13 R^\gamma_{\ \mu\beta\nu}z^\mu z^\nu\lambda^\beta
\end{equation}
and therefore the linear momentum and angular momentum generators are:

\begin{equation}
\label{4.4}
P^\gamma_\beta\approx\delta^\gamma_\beta-\frac13 R^\gamma_{\ \mu\beta\nu}z^\mu z^\nu,
\quad J^\alpha_{\rho\sigma}\approx x^\beta\eta_{\beta\gamma}\delta^{\alpha\gamma}_{\rho\sigma}
\end{equation}

Other natural systems of coordinates to use with a deeper physical meaning in the neighborhood of an event $x$ are those adapted to the isometry germs for which the space-time metric is independent of time up to the second order of approximation. More precisely we shall say that a system of coordinates {\bfseries y} is adapted to an isometry germ generated by $a: (\blambda, \bzeta)$ if in this system of coordinates we have:

\begin{equation}
\label{4.5}
\xi^0\approx 1, \quad \xi^i\approx 0, \quad i,j,k\cdots = 1,2,3
\end{equation}

To construct such system of coordinates requires to proceed with a two steps process: i) To perform a Lorentz transformation {\bfseries  L} to have:

\begin{equation}
\label{4.6}
\lambda^\alpha\sim \delta^\alpha_0
\end{equation}
and ii) to perform a coordinate transformation: 

\begin{equation}
\label{4.7}
y^\alpha\approx z^\alpha+\frac12 A^\alpha_{\rho\sigma}z^\rho z^\sigma
+\frac16 A^\alpha_{\rho\sigma\tau}z^\rho z^\sigma z^\tau
\end{equation}
to implement the conditions (\ref{4.5}). Assuming that the first step has already been made, this leads to the following equations that the completely symmetric {\bfseries A}'s have to satisfy:

\begin{eqnarray}
\label{4.8}
A^\alpha_{0\rho}&=&-\zeta_{\ \rho}^\alpha \\
\label{4.8.1}
A^\alpha_{0\rho\sigma}&=&-A^\alpha_{\lambda\sigma}\zeta_{\ \rho}^\lambda
-A^\alpha_{\lambda\rho}\zeta_{\ \sigma}^\lambda
+\frac13(R^\alpha_{\ \rho 0\sigma}+R^\alpha_{\ \sigma 0\rho})
\end{eqnarray} 
Since the {\bfseries A}'s below do not appear in the preceding equations we are free to impose the supplementary conditions: 

\begin{equation}
\label{4.9}
A^\alpha_{ij}=0, \quad A^\alpha_{ijk}=0
\end{equation}
From  Eqs. (\ref{4.13.2}) below we see that the choice above means that the space coordinates in the neighborhood of $x$ are geodesic coordinates of the space metric.

More explicitly the three indices {\bfseries A}'s are found to be:

\begin{eqnarray}
\label{4.11}
A^\alpha_{000}&=&2\zeta_{\ \beta}^\alpha\zeta_{\ 0}^\beta \\
A^\alpha_{00j}&=&-\frac13 R^\alpha_{\ 0j0}+
\zeta_{\ 0}^\alpha\zeta_{\ j}^0+
\zeta_{\ i}^\alpha\zeta_{\ j}^i  \\
A^\alpha_{0ij}&=&\frac13 (R^\alpha_{\ i0j}+R^\alpha_{\ j0i})+
\zeta_{\ i}^\alpha\zeta_{\ j}^0+
\zeta_{\ j}^\alpha\zeta_{\ i}^0 
\end{eqnarray}

To obtain the new expression for the space-time metric is a quite elementary but cumbersome process. It requires using (\ref{4.8})-(\ref{4.9}) to determine the coefficients of the inverse coordinate transformation:

\begin{equation}
\label{4.12}
z^\alpha\approx y^\alpha+\frac12 B^\alpha_{\rho\sigma}x^\rho x^\sigma+\frac16 B^\alpha_{\rho\sigma\tau}x^\rho x^\sigma x^\tau
\end{equation}
and going through the usual process of performing a tensor transformation on the metric (\ref{4.2}). The result is the following:

\begin{eqnarray}
\label{4.13}
g_{00}&\sim&-1+2\zeta_{0i}x^i+(\zeta_{\rho i} \zeta^\rho_j-R_{0i0j})x^ix^j\\
\label{4.13.1}
g_{0i}&\sim&-\zeta_{ji}x^j-\frac23 R_{ij0k}x^jx^k\\
\label{4.13.2}
g_{ij}&\sim&\delta_{ij}-\frac13 R_{ikjl}x^kx^l
\end{eqnarray}

Every time-like Killing congruence describes the motion of a frame of reference
and when using adapted coordinates the local gravitational field is described by the two space tensors:

\begin{equation}
\label{4.14}
\Lambda_i=-\partial_i \ln\xi, \quad \Omega_{ij}=\xi(\partial_i\varphi_j-\partial_j\varphi_i))
\end{equation}
with:

\begin{equation}
\label{4.15}
\xi=\sqrt{-g_{00}}, \quad \varphi_i=\xi^{-2}g_{0i}
\end{equation}
which are the Newtonian and Coriolis fields. 
We see from (\ref{4.13}) and (\ref{4.13.1}) that for isometry germs we have:

\begin{equation}
\label{4.16}
\Lambda_i\sim \zeta_{0i}, \quad \Omega_{ij}\sim 2\zeta_{ji}
\end{equation} 
showing that the elements $a: (\blambda,\bzeta)$ have a dual meaning since they are the generators of isometry germs as well as the local measurable gravitational fields in the frames of reference that they define.  

\section{Generalized Lorentz transformations: a novel structure}

Given two isometry germs $\bf \xi_1$ and $\bf \xi_2$ generated by $a_1: (\bf \lambda_1, \bf \zeta_1)$ and $a_2: (\bf \lambda_2, \bf \zeta_2)$ with the corresponding systems 
of adapted coordinates $\bf y_1$ and $\bf y_2$ obtained as in the preceding section, an important physical point is to establish the transformation from one of these systems of coordinates to the other. This is crucial to the very concept of physical relativity since these are the transformations that give an intrinsic  meaning to tensors, i.e. whose space-time physical quantities that exhibit different projections depending on the reference frame that different observers might use. If the Riemann tensor is zero and $\bf \zeta_1$ and $\bf \zeta_2$ are 
zero we know that these transformations are the Lorentz transformations. Therefore the important point we are referring to is how to determine the general transformations from ${\bf y}_1$ to ${\bf y}_2$, say, when i) the Riemann tensor is not zero at some event $x$ and/or ii) the isometry germs associated with two reference frames in the neighborhood of this event have general parameters ($\bf\lambda_1,\bf\zeta_1)$ and $(\bf\lambda_2,\bf\zeta_2)$.

Let $\bf L_1$ and $\bf L_2$ be two Lorentz transformations such that:

\begin{equation}
\label{5.1.1}
\delta^{\alpha^\prime}_{0^\prime}=L^{\alpha^\prime}_{1\rho}\lambda^\rho_1, \quad 
\delta^{\alpha^{\prime\prime}}_{0^{\prime\prime}}=L^{\alpha^{\prime\prime}}_{2\rho} \lambda^\rho_2
\end{equation}
then the transformations from ${\bf y}_1$ to ${\bf y}_2$ will have symbolically the following form:

\begin{equation}
\label{5.0}
{\bf y}_2 \approx f({\bf y}_1; {\balpha}_1; {\balpha}_2)
\end{equation} 
where each symbol $\balpha$ stands collectively for ($\bf L,\blambda,\bzeta$)
and as it can be easily seen by construction the function $f$ will have the following properties:
\begin{itemize}
\item 
\begin{equation}
f({\bf y};{\balpha};{\balpha})={\bf y}
\end{equation}
\item
\begin{equation}
f({\bf y}_2;{\balpha}_2;{\balpha}_1) \approx {\bf y_1}
\end{equation}
\item
\begin{equation}
f(f({\bf y}_1; {\balpha}_1; {\balpha}_2);{\balpha}_2;{\balpha}_3) \approx
f({\bf y}_1; {\balpha}_1; {\balpha}_3)
\end{equation}
\end{itemize}
These three properties define the novel structure that we mentioned in the abstract and the introduction\,\footnote{A similar structure was anticipated in \cite{Bel}}.

Again the details of the calculation in the general case are quite elementary but cumbersome and we shall restrict ourselves two consider two simple examples. The first one assumes: i) that the Riemann tensor is zero, or can be neglected:

\begin{equation}
\label{5.1}
R_{\alpha\beta\gamma\delta}\sim 0
\end{equation}
and ii) that the product of $\bzeta$'s can be neglected. Using the self-explanatory notations:

\begin{equation}
\label{5.1.0}
z_1^{\alpha^\prime}=L^{\alpha^\prime}_{1\rho}z^\rho, \quad 
z_2^{\alpha^{\prime\prime}}=L^{\alpha^{\prime\prime}}_{2\rho}z^\rho
\end{equation}
the two coordinate transformations (\ref{4.7}) to be considered are:

\begin{eqnarray}
\label{5.2}
y_2^{\alpha^{\prime\prime}} &\approx & z_2^{\alpha^{\prime\prime}} +\frac12 A^{\alpha^{\prime\prime}}_{2\mu^{\prime\prime}\nu^{\prime\prime}}
z_2^{\mu^{\prime\prime}}z_2^{\nu^{\prime\prime}} \\
\label{5.2.2.0}
y_1^{\alpha^\prime} &\approx & z_1^{\alpha^\prime} +
\frac12 A^{\alpha^\prime}_{1\mu^\prime\nu^\prime}
z_1^{\mu^\prime}z_1^{\nu^\prime}
\end{eqnarray}    
and its inverse at the corresponding approximation, where the non zero $\bf A$'s are given by Eqs. (\ref{4.8}) for both sets of indices with one prime or two primes.

Eliminating $\bf z$ from (\ref{5.2}) using the inverse of (\ref{5.2.2.0}) we get:

\begin{equation}
\label{5.3.1}
y_2^{\alpha^{\prime\prime}}\approx L^{\alpha^{\prime\prime}}_{3\gamma\prime}
(y_1^{\gamma^\prime} -\frac12
A^{\gamma^\prime}_{1\mu^\prime\nu^\prime}y_1^{\mu^\prime}y_1^{\nu^\prime})
+\frac12 A^{\alpha^{\prime\prime}}_{2\mu^{\prime\prime}\nu^{\prime\prime}}
L^{\mu^{\prime\prime}}_{3\gamma^\prime}
L^{\nu^{\prime\prime}}_{3\delta^\prime}y^{\gamma^\prime}y^{\delta^\prime} 
\end{equation}
where:

\begin{equation}
\label{5.3.1.1}
L^{\alpha^{\prime\prime}}_{3\gamma\prime}= 
L^{\alpha^{\prime\prime}}_{2\rho}L^\rho_{1\gamma\prime}
\end{equation}
If in particular we assume that ${\bf L}_2=\bdelta$ then we have dropping the double primes:

\begin{equation}
\label{5.4}
y_2^\alpha\approx L^\alpha_{1\gamma\prime}
(y_1^{\gamma^\prime} -\frac12
A^{\gamma^\prime}_{1\mu^\prime\nu^\prime}y_1^{\mu^\prime}y_1^{\nu^\prime})
+\frac12 A^\alpha_{2\mu\nu}
L^\mu_{1\gamma^\prime}
L^\nu_{1\delta^\prime}y_1^{\gamma^\prime}y_1^{\delta^\prime}
\end{equation}
or, using (\ref{4.8}):

\begin{eqnarray}
\label{5.4.0}
y_2^\alpha\approx L^\alpha_{1\gamma\prime}
(y_1^{\gamma^\prime} 
-\frac12\zeta_{10^\prime}^{\gamma^\prime}(y_1^{0^\prime})^2 
-\zeta_{1j^\prime}^{\gamma^\prime}y_1^{0^\prime} y_1^{j^\prime} \nonumber \\
+\frac12(\zeta_{10}^{\alpha}L^0_{1\gamma^\prime}L^0_{1\delta^\prime}
+\zeta_{1j}^\alpha
(L^0_{1\gamma^\prime}L^j_{1\delta^\prime}+L^0_{1\delta^\prime}L^j_{1\gamma^\prime}))
y_1^{\gamma^\prime}y_1^{\delta^\prime}
\end{eqnarray}

And if we add the assumption $\bf L_2=\bdelta$ then we have
dropping the tildes:

\begin{equation}
\label{5.4.1}
y_2^\alpha \approx y_1^\alpha 
-\frac12(\zeta_{10}^\alpha-\zeta_{20}^\alpha)(y_1^0)^2 -(\zeta_{1j}^\alpha-\zeta_{2j}^\alpha)y_1^0 y_1^j
\end{equation}

In the second example we assume that both $\bzeta_1$ and $\bzeta_2$ are zero. The two coordinate transformations (\ref{4.7}) that we have to use now are:

\begin{eqnarray}
\label{5.2.0}
y_2^{\alpha^{\prime\prime}} &\approx & z_2^{\alpha^{\prime\prime}} +\frac16 A^{\alpha^{\prime\prime}}_{2\mu^{\prime\prime}\nu^{\prime\prime}\rho^{\prime\prime}}
z_2^{\mu^{\prime\prime}}z_2^{\nu^{\prime\prime}}z_2^{\rho^{\prime\prime}} \\
\label{5.2.0.1}
y_1^{\alpha^\prime} &\approx & z_1^{\alpha^\prime} +
\frac16 A^{\alpha^\prime}_{1\mu^\prime\nu^\prime\rho^\prime}
z_1^{\mu^\prime}z_1^{\nu^\prime}z_1^{\rho^\prime}
\end{eqnarray}    
and its inverse at the corresponding approximation, where the non zero $\bf A$'s are:

\begin{equation}
\label{5.2.2}
A^\alpha_{00j}=R^\alpha_{\ 00j}, \quad A^\alpha_{0ij}=
\frac12(R^\alpha_{\ i0j}+R^\alpha_{\ j0i}) 
\end{equation}
for both $\bf A$'s with indices with one prime or two primes.
Eliminating $\bf z$ from (\ref{5.2.0}) using the inverse (\ref{5.2.0.1}) we get:

\begin{equation}
\label{5.3.1.2}
y_2^{\alpha^{\prime\prime}}\approx L^{\alpha^{\prime\prime}}_{3\gamma\prime}
(y_1^{\gamma^\prime} -\frac16
A^{\gamma^\prime}_{1\mu^\prime\nu^\prime\rho^\prime}y_1^{\mu^\prime}y_1^{\nu^\prime}y_1^{\rho^\prime})
+\frac16 A^{\alpha^{\prime\prime}}_{2\mu^{\prime\prime}\nu^{\prime\prime}\rho^\prime}
L^{\mu^{\prime\prime}}_{3\gamma^\prime}
L^{\nu^{\prime\prime}}_{3\delta^\prime}L^{\rho^{\prime\prime}}_{3\epsilon^\prime}
y_1^{\gamma^\prime}y_1^{\delta^\prime}y_1^{\epsilon^\prime} 
\end{equation}
If we consider now the case with $\bf L_2=\bdelta$ then we have dropping the double primes:

\begin{equation}
\label{5.5}
y_2^\alpha\approx L^\alpha_{1\gamma\prime}
(y_1^{\gamma^\prime} -\frac12
A^{\gamma^\prime}_{1\mu^\prime\nu^\prime\rho^\prime}y_1^{\mu^\prime}y_1^{\nu^\prime}y_1^{\nu^\prime})
+\frac12 A^\alpha_{2\mu\nu\rho}
L^\mu_{1\gamma^\prime}
L^\nu_{1\delta^\prime}L^\rho_{1\epsilon^\prime} y_1^{\gamma^\prime}y_1^{\delta^\prime}y_1^{\epsilon^\prime}
\end{equation}
or using (\ref{5.2.2}):

\begin{eqnarray}
\label{5.6}
y_2^{\alpha^\prime}\approx
L^\alpha_{\gamma^\prime}
(y_1^\rho-\frac16 R^{\gamma^\prime}_{\ 0^\prime 0^\prime j^\prime} (y_1^{0^\prime})^2 y_1^{0^\prime}y_1^{j^\prime}
-\frac13 R^{\gamma^\prime}_{\ i^\prime 0^\prime j^\prime}
y_1^{i^\prime}y_1^{0^\prime}y_1^{j^\prime}) \nonumber \\ 
+\frac13(\frac12 R^\gamma_{\ 00j}
L^0_{1\gamma^\prime}L^0_{1\delta^\prime} L^j_{1\epsilon^\prime} 
+R^\gamma_{\ i0j}L^0_{1\gamma^\prime} 
L^i_{1\delta^\prime}
L^j_{1\epsilon^\prime})y_1^{\gamma^\prime} y_1^{\delta^\prime} y_1^{\epsilon^\prime}
\end{eqnarray}

If we further assume that $\bf L_1=\bdelta$ then we would get $\bf y_2=\bf y_1$.

\section{Acknowledgments} 

I gratefully acknowledge the hospitality of the Theoretical Physics Department of the UPV/EHU as well as useful information and comments from Mart\'{\i}n Rivas.


\begin{thebibliography}{9}

\bibitem{Lichnerowicz} A.\ Lichnerowicz, \textsl{Geom$\acute{e}$trie des Groupes de Transformations}, Chap. II,  Sect. 30, Dunod (Paris 1958)

\bibitem{Yano} K.\ Yano, \textsl{Theory of Lie Derivatives and its Applications}, North Holland (1955)

\bibitem{Eisenhart} L.\ P.\ Eisenhart, \textsl{Riemannian Geometry} Chap. VI Sect. 70-71, Princeton University Press (1949)

\bibitem{Dixmier} J.\ Dixmier \textsl{Alg$\grave{e}$bres de Lie}, Chap. I Sect. 10, Chap. III Sect. 2, Centre de Documentation Universitaire (Paris)

\bibitem{Hammermesh} M.\ Hammermesh, \textsl{Group Theory}, Addison-Wesley (1964)

\bibitem{Gursey} F.\ G\"{u}rsey, in \textsl{Relativity Groups \& Topology}, Sects. IV and V, Ed. by C.\ DeWitt and B.\ DeWitt, Blackie and Son Limited (1964)

\bibitem{Okubo} S.\ Okubo, \textsl{J.\ Math.\ Phys.}, {\bf 18}, 2382 (1977)

\bibitem{Thomas} T.\ Y.\ Thomas, \textsl{Concepts from Tensor Analysis and Differential Geometry}, Academic Press (New York 1961)

\bibitem{Bel} Ll. Bel, in \textsl{Relativity and Gravitation in
General}, edited by J. Mart\'{i}n, E. Ruiz, F. Atrio and A. Molina.
World Scientific. Also gr-qc/9812062

\end{thebibliography}
\end{document}